\documentstyle[prl,aps,twocolumn,epsf]{revtex}
\def\break#1{\pagebreak \vspace*{#1}}
\begin{document}

%\draft

\title{Ermakov-Lewis angles for one-parameter supersymmetric families of
Newtonian free damping modes}

\author{Haret C. Rosu%\footnote{Electronic mail: rosu@ifug3.ugto.mx}
$^{\dagger}$%$^{\ddagger}$
%\cite{byline}
and Pedro B. Espinoza%\footnote{Electronic mail:pedroe@ifug3.ugto.mx}
$^{\ddagger}$
}

\address{ $^{\dagger}$
Instituto de F\'{\i}sica de la Universidad de Guanajuato, Apdo Postal
E-143, Le\'on, Guanajuato, M\'exico and}
\address{International Center for Relativistic Astrophysics, Rome-Pescara, Italy}
\address{$^{\ddagger}$
Centro Universitario de los Altos, Universidad de Guadalajara, Lagos de
Moreno, Jalisco, M\'exico
 }

%\date{submitted to PRe}
\maketitle
\widetext

\begin{abstract}
We apply the Ermakov-Lewis procedure to the one-parameter
damped modes $\tilde{y}$ recently introduced by Rosu and Reyes, which
are related to the common Newtonian free damping modes $y$ by the general 
Riccati solution
[H.C. Rosu and M. Reyes, Phys. Rev. E {\bf 57}, 4850 (1998)].
In particular, we calculate and plot the angle quantities of this approach
that can help to
distinguish these modes from the common $y$ modes.
\end{abstract}
\vskip 0.1in

PACS number(s):  03.20.+i
\vskip 0.1in

%\pacs{PACS numbers: 98.80.Hw, 11.30Pb, 04.60.Kz} %\hspace{7cm}
%LAA number: physics/0004xxx}

%%%%%%%%%%%%%%%%%%%%%%%%        THE PAPER       %%%%%%%%%%%%%%%%%%%%%%%%
%%%%%%%%%%%%%%%%%%%%%%%%%  written by H.C. Rosu  %%%%%%%%%%%%%%%%%%%%%%%%%%%
%%%%%%%%%%%%%%%%%%%%%%%%       April 2000       %%%%%%%%%%%%%%%%%%%%%%
\narrowtext

%%%%%%%%%%%%%%%%%%%%%%%%%%%%%%%%%%%%%%%%%%%%%%%%%%%%%%%%%%%%%%%%%%%%%%%
%{\bf 1}.-
%particular solutions of Riccati equations known as superpotentials,
%we would like to explore here the factoring of the DO
%equation by means of the general solution of the Riccati equation, a
%procedure which has been first used in physics by Mielnik \cite{M} for the
%quantum harmonic oscillator.
%%%%%%%%%%%%%%%%%%%%%%%%%%%%%%%%%%%%%%%%%%%%%%%%%%%%%%%%%%%%
In a previous paper hereafter denoted as I \cite{rr98}, the
non-uniqueness of the factorization of linear second-order differential 
operators has been exploited
on the example of the classical Newtonian free damped oscillator, i.e.
%%%%%%%%%%%%%%
$$
Ny\equiv
\left(\frac{d^2}{dt^2}+2\beta\frac{d}{dt}+\omega _{0}^{2}\right)y=0~.
\eqno(1)
$$
%%%%%%%%%%%%%%%%
The coefficient $2\beta$ is the friction constant per unit
mass and
%\break{0.2in}
$\omega _{0}$ is the natural frequency of the oscillator (SI units assumed all
over the work).

The more general supersymmetric partner equation
%%%%%%%%%%%%%%%%%%%%%
$$
\tilde{N}_{g}\tilde{y}\equiv
\left(\frac{d^2}{dt^2}+2\beta\frac{d}{dt}+\omega _{0} ^{2}-
\frac{2\gamma ^2}{(\gamma t+1)^2}\right)\tilde{y}=0
\eqno(2)
$$
%%%%%%%%%%%%%%%%%%%%%%%%
has been obtained in I.
This new second-order linear damping equation contains the additional
last term with
respect to its initial partner (1), which may be thought of as the general
Darboux transform part of the frequency \cite{D}.
$T=1/\gamma$ occurs as a new time scale in the Newtonian damping problem.
If this time scale is infinite, the ordinary free damping is recovered 
unless for the critical case which is special even in ordinary damping.
As explained in I, the $\tilde{y}$ modes can be obtained from the $y$ modes
by operatorial means. In the following we shall call them $\gamma$ modes.
For the three types of
free damping they have been obtained in I as follows:

(i) For underdamping, $\beta ^{2}<\omega _{0}^{2}$, denoting
%$\alpha =i\omega _1$, where
$\omega _{u}=\sqrt{\omega _{0}^{2}-\beta ^2}$ 
%The original eigenfunction is
%$y_{u}=\tilde{A}_{u}\cos(\omega _{1}t+\phi)e^{-\beta t}$,
the underdamped $\gamma$ modes are  
$$
\tilde{y} _{u}= -\tilde{A} _{u}e^{-\beta t}
\Big[\omega _{u}\sin(\omega _{u}t+\phi)+\frac{\gamma}{\gamma t+1}
\cos(\omega _{u}t+\phi)\Big]~.
\eqno(3)
$$

%\break{0.98in}
(ii) For overdamping, $\beta ^2>\omega _{0}^{2}$ and 
$\omega _{o}=\sqrt{\beta ^2-\omega _{0}^{2}}$,
%the initial free
%general solution is $y_{o}=\tilde{A} _{o}e^{-\beta t}\cosh(\alpha t+\phi)$,
%A_{o,\pm}e^{-\beta t\pm\alpha t}$,
the overdamped $\gamma$ modes are
%eigenfunctions read
%$\tilde{y}_{o,\pm}=A_{o,\pm}(\pm \alpha -\frac{\gamma}{\gamma t+1})
%e^{-\beta t\pm \alpha t}$.
$$
\tilde{y} _{0}=-\tilde{A} _{o}e^{-\beta t}\Big[\omega _{o} 
\sinh(\omega _{o} t+\phi)-
\frac{\gamma}{\gamma t +1}\cosh (\omega _{o} t+\phi)\Big]~.
\eqno(4)
$$

(iii) For critical damping, $\beta ^2=\omega _{0}^{2}$.
%the general free solution is
%$y_{c}=Ae^{-\beta t}+Bte^{-\beta t}$, whereas
The critical $\gamma$ solutions are given by
$$
\tilde{y}_{c}=\Big[\frac{-A_{c}\gamma}{\gamma t+1}+\frac{D_{c}}{\gamma ^2}
(\gamma t +1)^2\Big]e^{-\beta t}~.
\eqno(5)
$$

\break{0.98in}

These are the {\em only} possible types of one-parameter
damping modes related to
the free damping ones by means of Witten's supersymmetric scheme \cite{Wi}
and the general Riccati solution \cite{M}.

In practice the new parameter $\gamma$ can be very close to zero.
In this case, it is very difficult to differentiate the $\gamma$ modes 
from the ordinary
ones. The only means we can think of is by recording somehow
the geometric angle associated to the $\gamma$ modes and compare it with the same
quantity in the ordinary damping cases. One is led to this
conclusion noticing that the $\gamma$ modes have time-dependent frequencies
$\omega ^{2}(t)=\omega _{0} ^{2}-
\frac{2\gamma ^2}{(\gamma t+1)^2}$
and hence for them the Ermakov-Lewis (EL) procedure can be naturally applied
\cite{EL} (for a recent review, see \cite{Esp}).
%This work is devoted to the EL analysis of the $\gamma$ modes.
%Before proceeding, we make the following digression. 
For $\omega _{0}\neq \beta$, Eq.~(2) can be reduced to
a Bessel equation and the solutions can be written as follows
$$
\Psi _{u}=\tau ^{1/2}\Big[AJ_{\frac{3}{2}}(k\tau)+BY_{\frac{3}{2}}(k\tau)\Big]
e^{-\beta \tau}
\eqno(6)
$$
and
$$
\Psi _{o}=\tau ^{1/2}\Big[CI_{\frac{3}{2}}(k\tau)+DK_{\frac{3}{2}}(k\tau)\Big]
e^{-\beta \tau}~,
\eqno(7)
$$
where $\tau=\gamma t+1$ and 
$k^2=\frac{\omega _{0}^2-\beta ^2}{\gamma ^2}$. When
$k\rightarrow \infty$ (i.e., $\gamma \rightarrow 0$), we can do
Hankel's asymptotic expansions, i.e., of large Bessel argument but fixed
Bessel order (we shall not reproduce these formulas here,
the reader is directed to Abramowitz and Stegun \cite{AS}). The point is
that one is indeed able to get the solutions obtained by operatorial means from
inspecting Hankel's expansions.
Thus, the supersymmetric operatorial procedure gives merely the
asymptotic $\gamma \rightarrow 0$ solutions, which however could be the
most relevant from the physical viewpoint in this context.

In the EL approach the angular quantities are given by the following formulas
\cite{Esp,mor}
%%%%%%%%%%%%%
$$
\Delta \theta ^{{\rm d}}=
%\int _{0}^{T}
%\langle\frac{\partial H_{\rm{new}}}{\partial {\cal I}}\rangle
%d\Omega ^{'}=
\int _{0}^{T}\Big[\frac{e^{-2\beta t '}}{\rho ^2}-
\frac{1}{2}\frac{d}{dt ^{'}}
(e^{2\beta t ^{'}}\dot{\rho}\rho)+e^{2\beta t ^{'}}
\dot{\rho}^2\Big]d t ^{'}
\eqno(8)
$$
and
$$
\Delta \theta ^{{\rm g}}=\frac{1}{2}\int _{0}^{T}
\Big[\frac{d}{d t ^{'}}
(e^{2\beta t ^{'}}\dot{\rho}\rho)-2e^{2\beta t ^{'}}
\dot{\rho}^2\Big]
dt ^{'},
\eqno(9)
$$ 
for the dynamical and geometrical angles, respectively.
%%%%%%%%%%%%%%%
Thus, the total angle will be
%%%%%%%%%%%%%%%%%%%%%%
$$
\Delta \theta ^{{\rm t}} =\int _{0}^{T}\frac{e^{-2\beta t ^{'}}}
{\rho ^2}
dt ^{'}~.
\eqno(10)
$$
The so-called Pinney function $\rho$ is the solution of Pinney's 
nonlinear equation \cite{P}
$$
\rho''(t)+p(t)\rho'(t)+q(t)\rho=\frac{C}{\rho^3(t)}\exp\left(-2\int ^{t} p(t')dt'
\right)
\eqno(11)
$$
for $C=$ constant (=1), $p(t)=2\beta$ and $q(t)=\omega _{0}^{2}-
\frac{2\gamma ^2}{(\gamma t+1)^2}$.
%For the oscillator of constant frequency $\omega _{0}$ the Pinney function
%is $\rho =1$ implying a zero geometric angle. 
For $\rho\neq$ constant
there is a definite prescription of calculating $\rho$ in terms of two 
independent solutions of the corresponding linear equation. 
We have followed the method of Eliezer and Gray \cite{EG}
for $\rho(t)$ in terms of linear combinations 
of the aforementioned Bessel functions (for $A=B=C=D=1$) that satisfy
the initial conditions as given by those authors.
In the critical damping case, we used the modes of Eq.~(5) with $A_{c}=D_{c}=1$.
The results of the calculations for some particular values of the parameters
are plotted in Figs. 1a,b,c, 2a,b,c, 3a,b,c
for the $\gamma$ underdamped, overdamped, and critical cases, respectively.
For comparison, the angle quantities for $\gamma =0$, within the same
calculational scheme, are displayed  
in Figs. 1a',b',c', 2a',b',c', 3a',b',c', respectfully.

%%%%%%%%%%%%%%%%%%%%%%%%%%%%%%%%%%%%%%%%%%%%%%%%%%%%%%%%%%%%%%%%%%%%%
\section*{Acknowledgment}
This work was partially supported by the CONACyT Project 458100-5-25844E.

%%%%%%%%%%%%%%%%%%%%%%%%%%%%%%%%%%%%%%%%%%%%%%%%%%%%%%%%%%%%%%%%%%%%%%

%%%%%%%%%%%%%%%%%%%%%%%%%%%%%%%%%%%%%%%%%%%%%%%%%%%%%%%%%%%%%%%%%%%%%%%%%%

\newpage

\vskip 1ex
\centerline{
\epsfxsize=220pt
\epsfbox{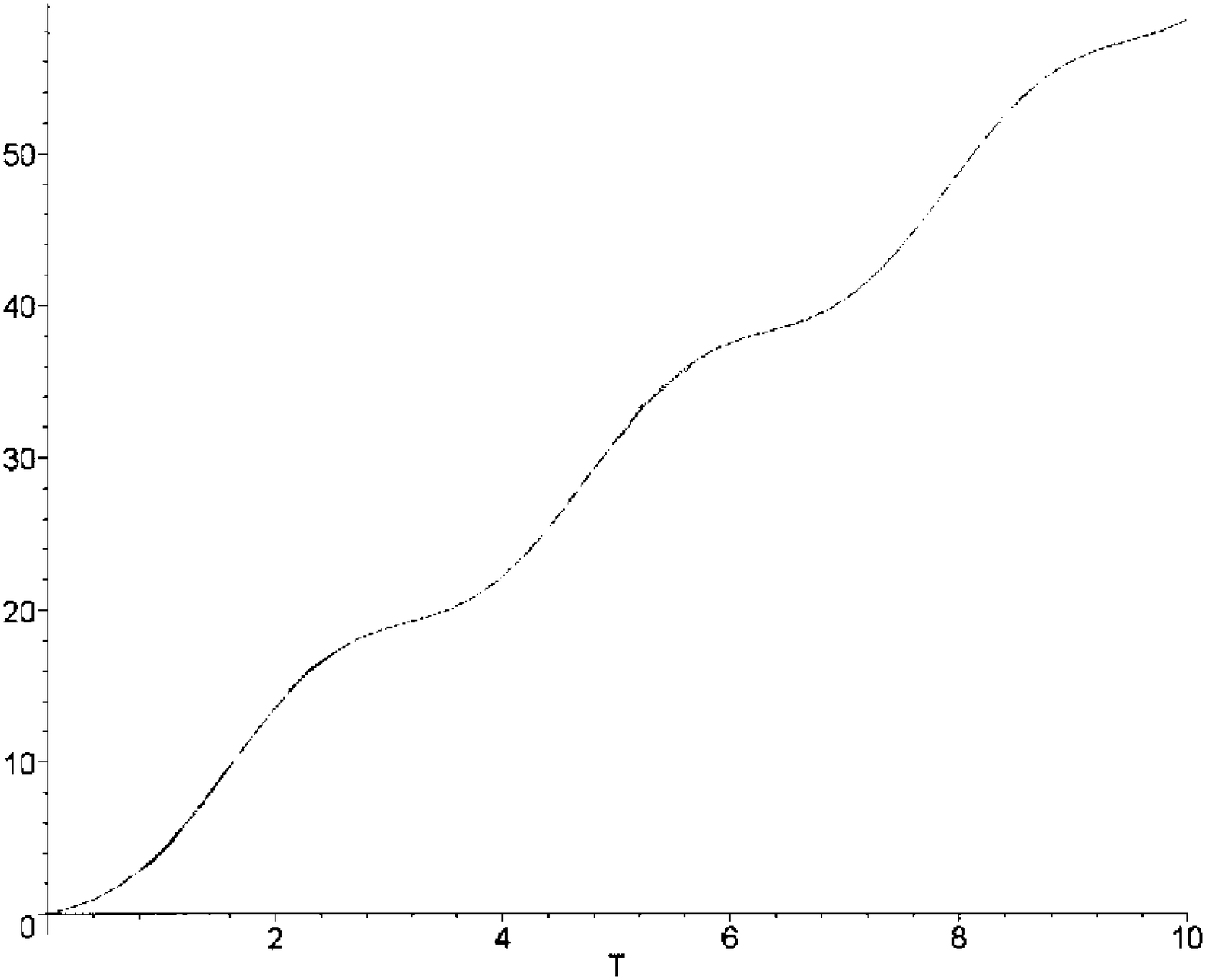}}
\vskip 3ex
\begin{center}
{\small{Fig. 1a}\\
 The dynamical angle in the underdamped case for the following set of
parameters: $\omega _{0}=\sqrt{2}$, $\beta=1$, $\gamma=0.1$. 
}
\end{center}

\vskip 1ex
\centerline{
\epsfxsize=220pt
\epsfbox{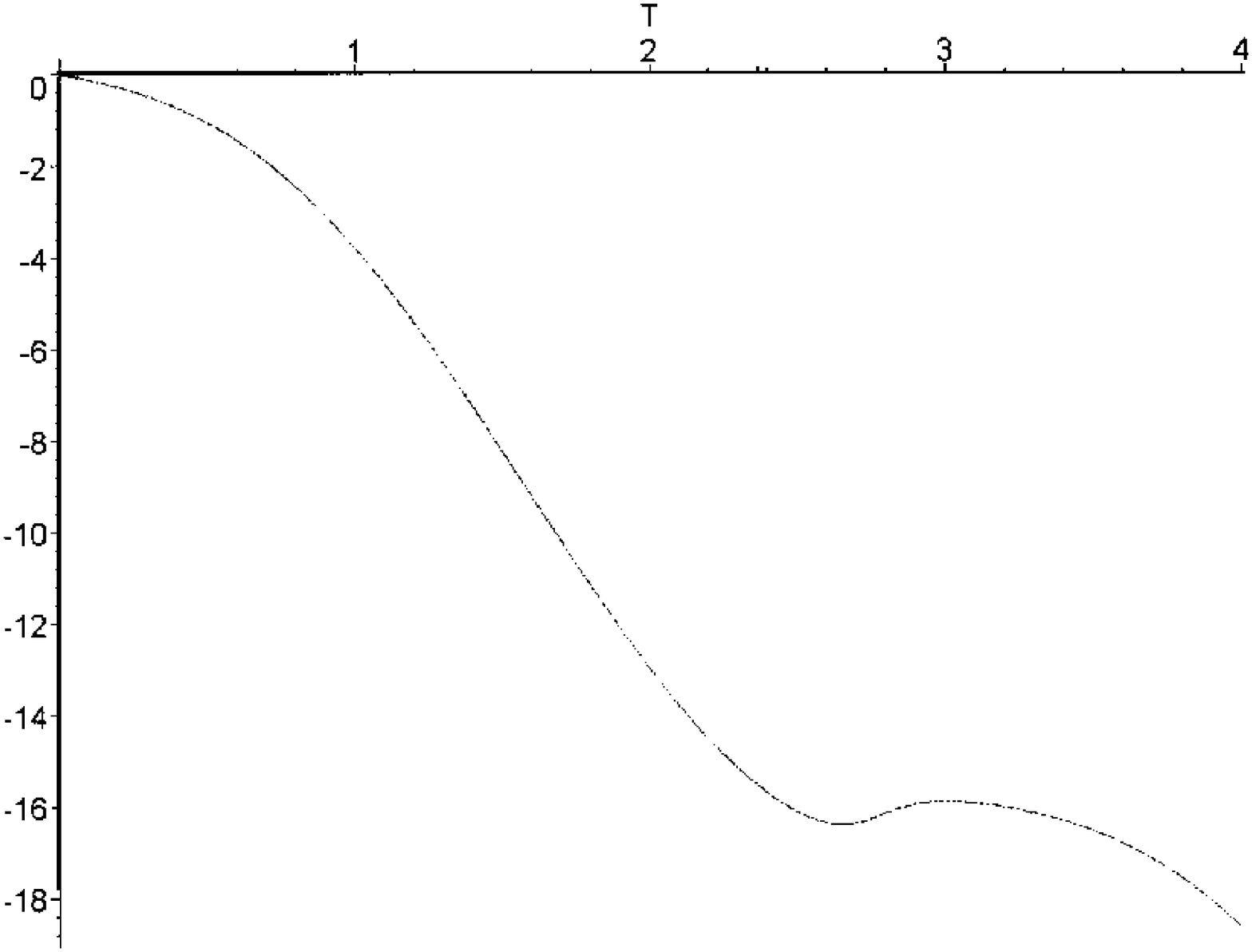}}
\vskip 3ex
\begin{center}
{\small{Fig. 1b}\\
 The geometric angle in the underdamped case and the same parameters. 
}
\end{center}

\vskip 1ex
\centerline{
\epsfxsize=220pt
\epsfbox{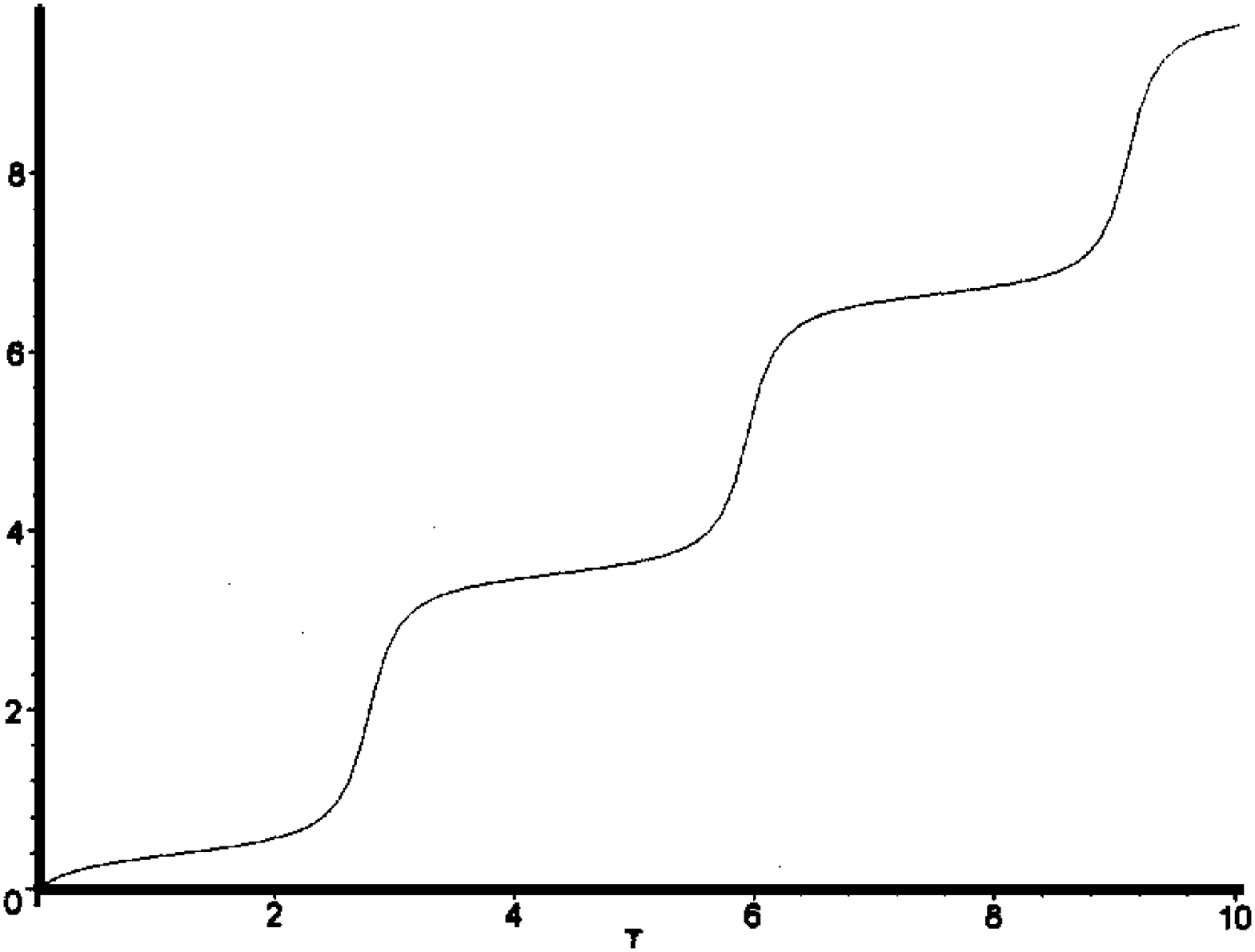}}
\vskip 3ex
\begin{center}
{\small{Fig. 1c}\\
 The total angle in the underdamped case and the same parameters. 
}
\end{center}

%%%%%%%%%%%%%%%%%%%%%%  end underdamping  %%%%%%%%%%%%%%%%%%%%%%%%%%%

\centerline{
\epsfxsize=200pt
\epsfbox{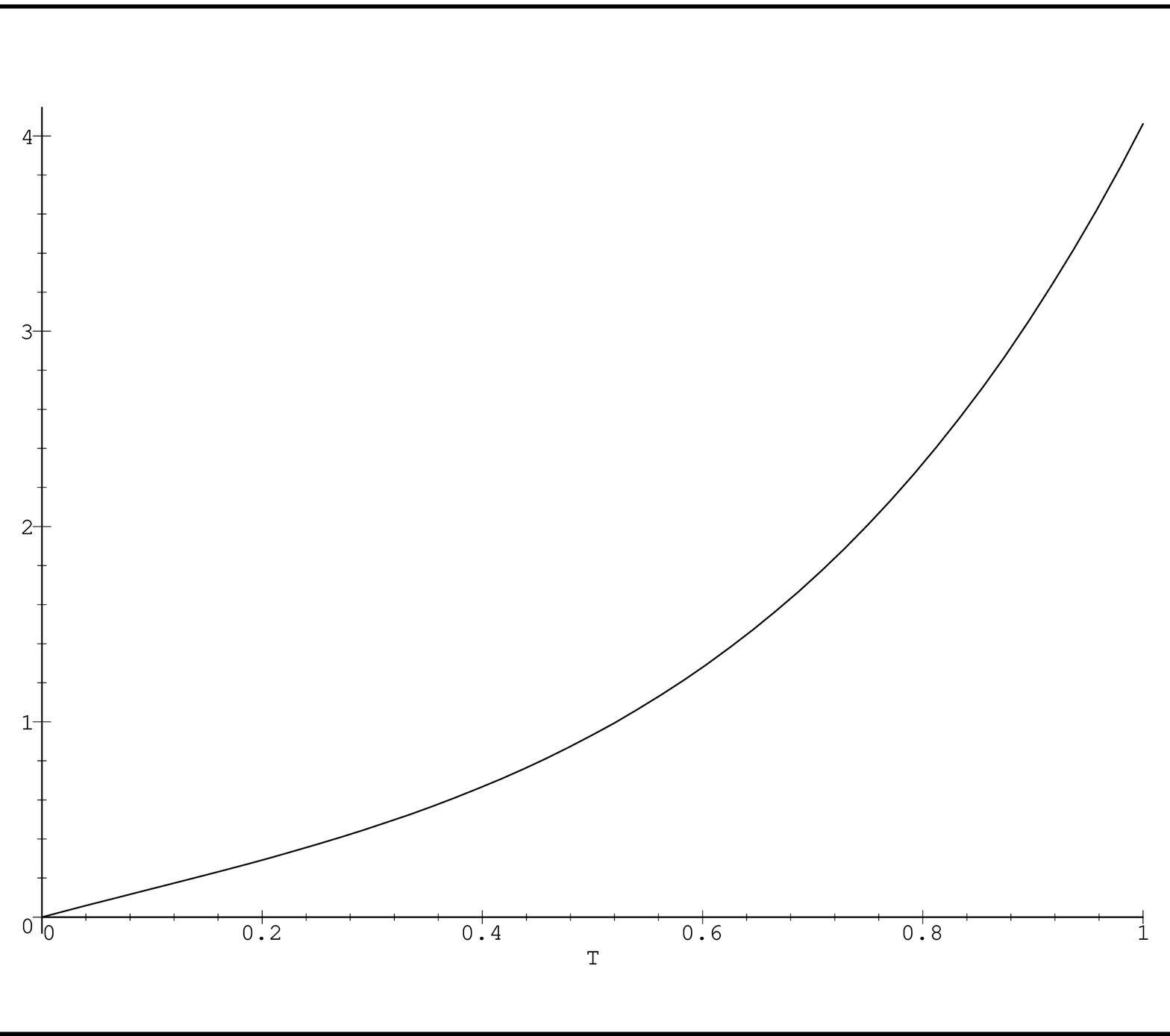}}   %ado1root.ps adchirpo11.ps
\vskip 3ex
\begin{center}
{\small{Fig. 2a}\\
The dynamical angle in the overdamped case for 
$\omega _{0}=1$, $\beta=\sqrt{2}$, $\gamma=0.1$. 
}
\end{center}

\vskip 1ex
\centerline{
\epsfxsize=200pt
\epsfbox{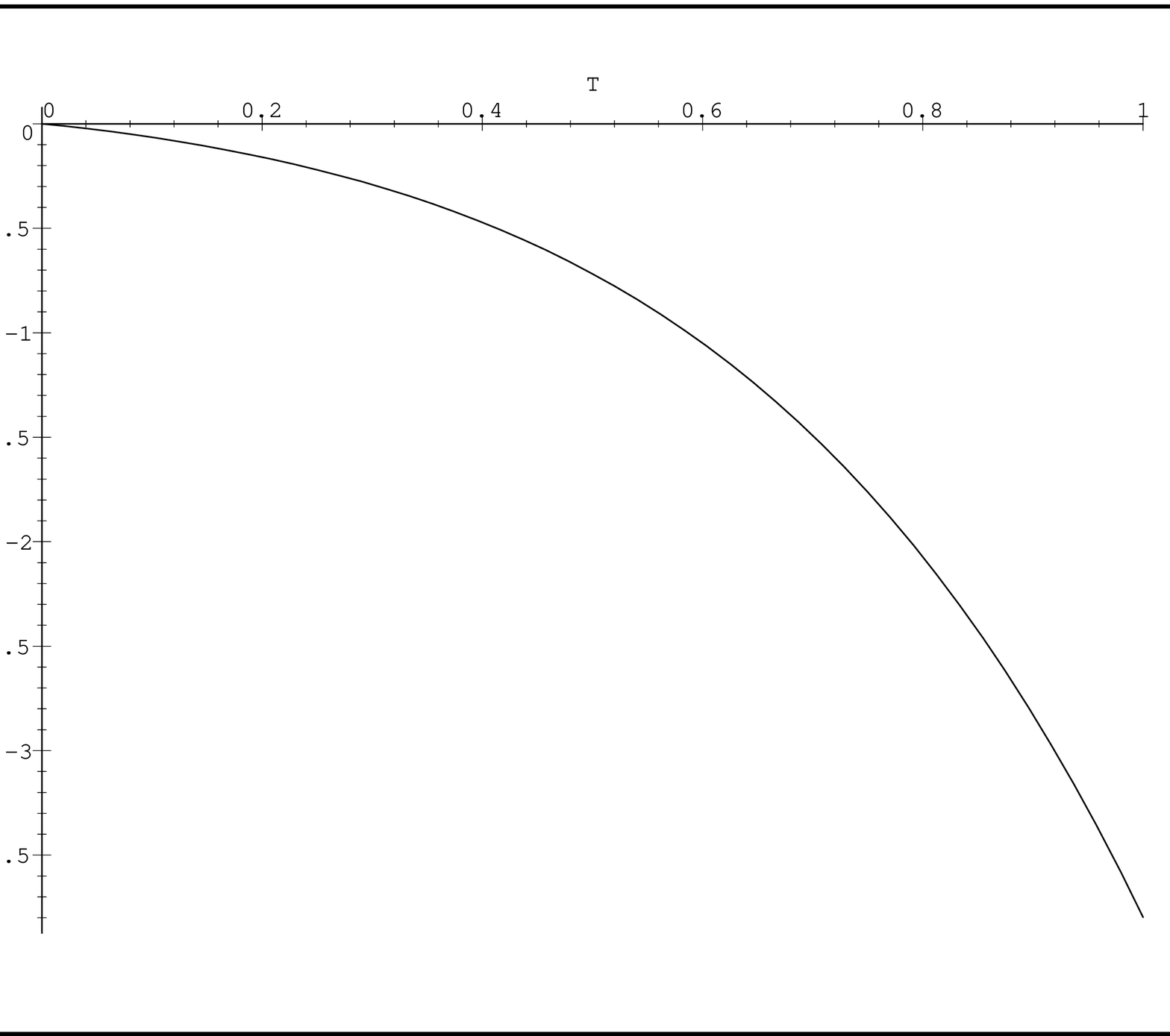}} %agchirpo1.ps  April 12
\vskip 3ex
\begin{center}
{\small{Fig. 2b}\\
 The geometric angle in the overdamping case for the same parameters.
}
\end{center}

\vskip 1ex
\centerline{
\epsfxsize=200pt
\epsfbox{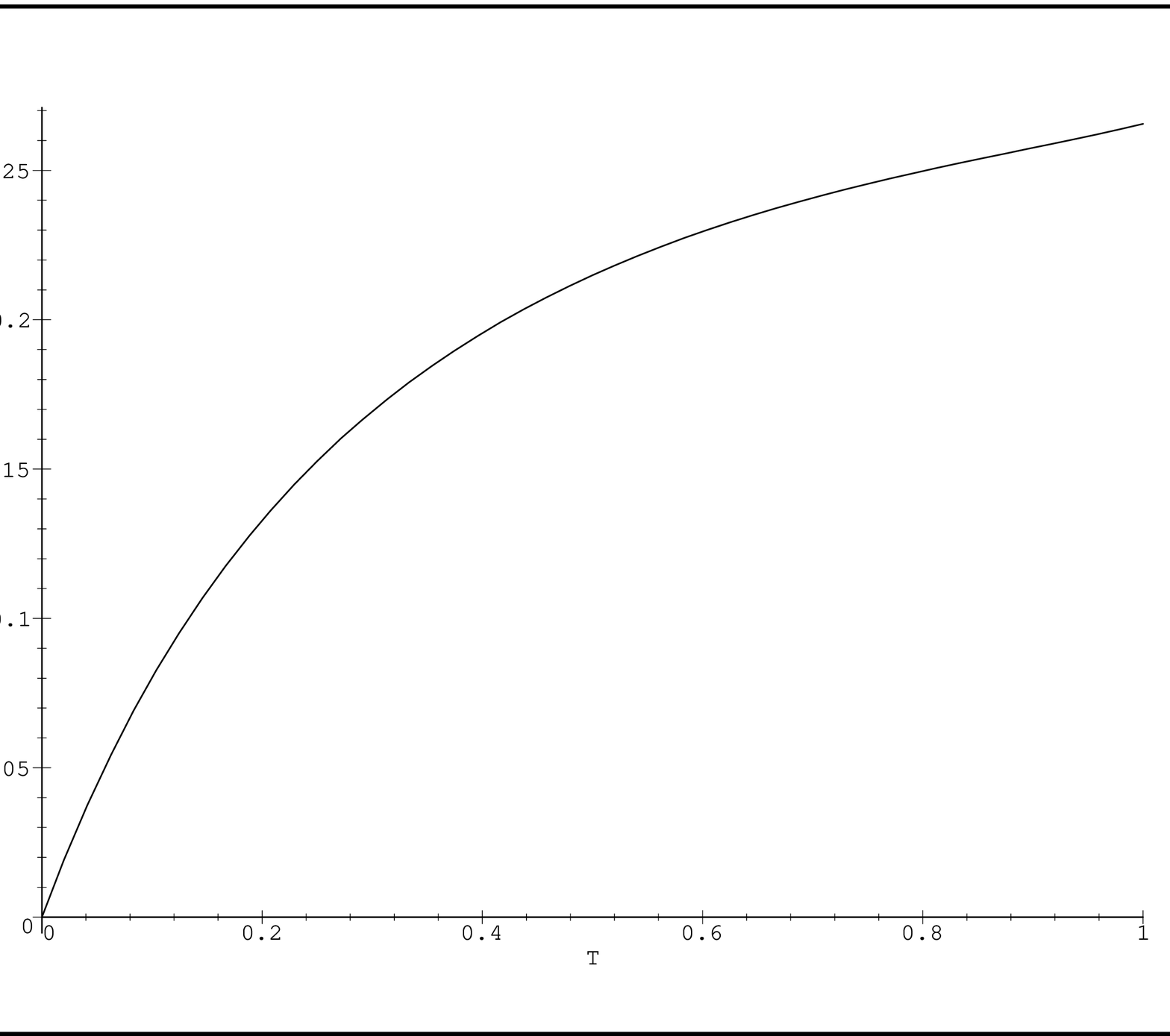}}   %atoroot.ps
\vskip 3ex
\begin{center}
{\small{Fig. 2c}\\
 The total angle in the overdamping case for the same parameters.
}
\end{center}

%%%%%%%%%%%%%%%%%%%%%%%%  end overdamping  %%%%%%%%%%%%%%%%%%%%%%%%%%%%%%%%

\vskip 1ex
\centerline{
\epsfxsize=200pt
\epsfbox{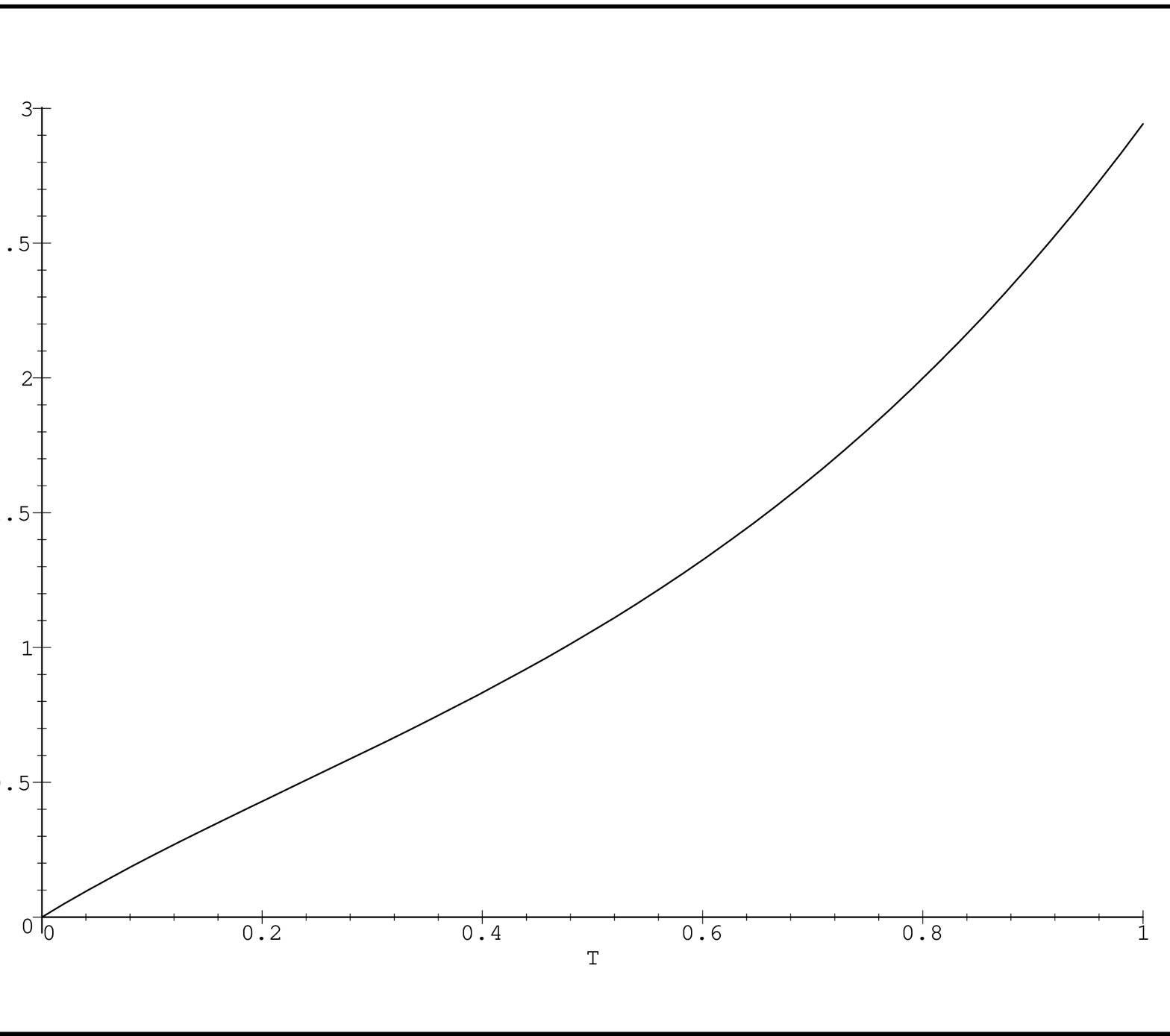}}
\vskip 3ex
\begin{center}
{\small{Fig. 3a}\\
 The dynamical angle in the critical case for $\omega _{0}=\beta=1$ and 
$\gamma=0.1$.  
}
\end{center}

\vskip 1ex
\centerline{
\epsfxsize=220pt
\epsfbox{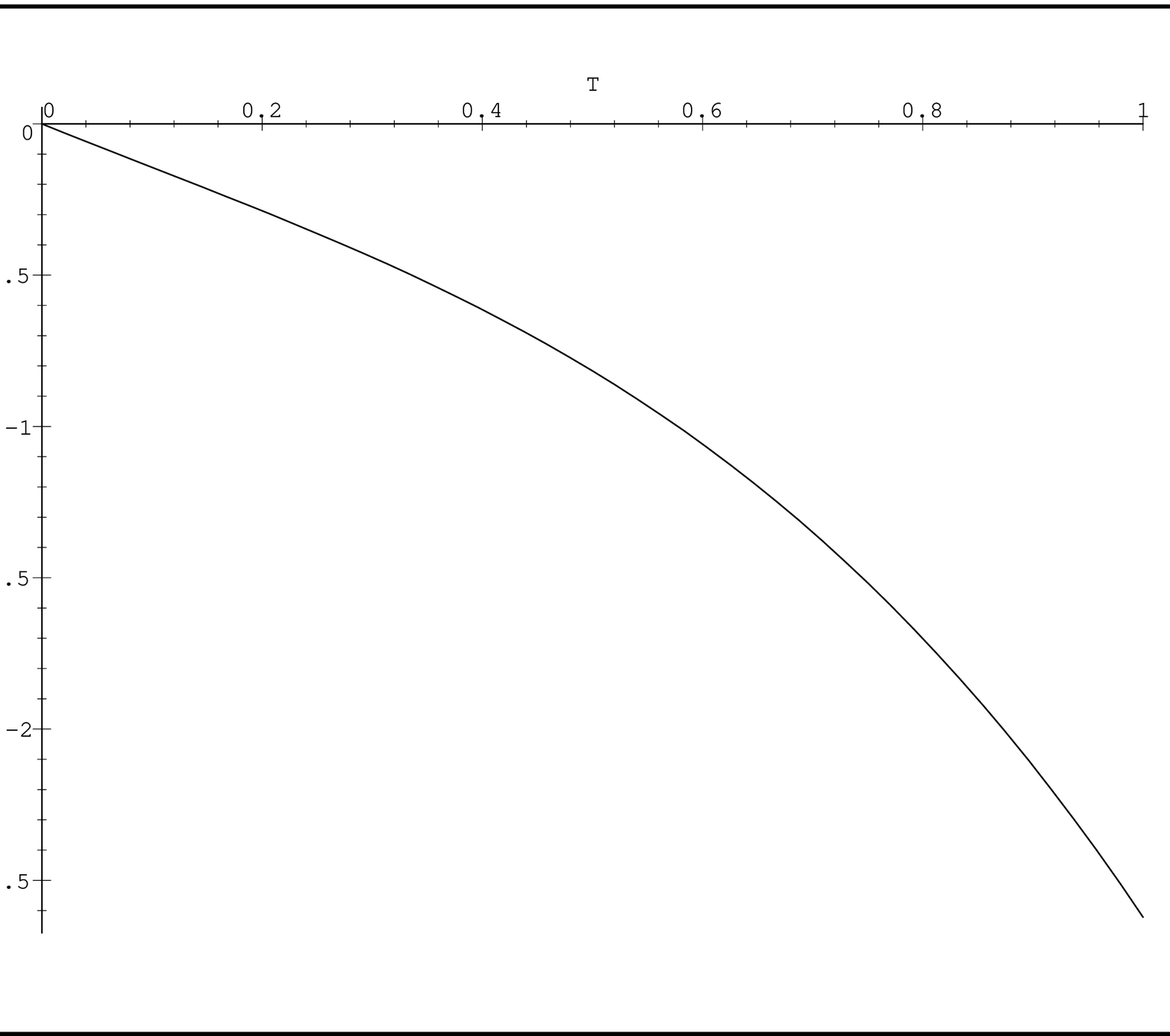}}
\vskip 3ex
\begin{center}
{\small{Fig. 3b}\\
The geometrical angle in the critical case for the same parameters.
}
\end{center}

\vskip 1ex
\centerline{
\epsfxsize=200pt
\epsfbox{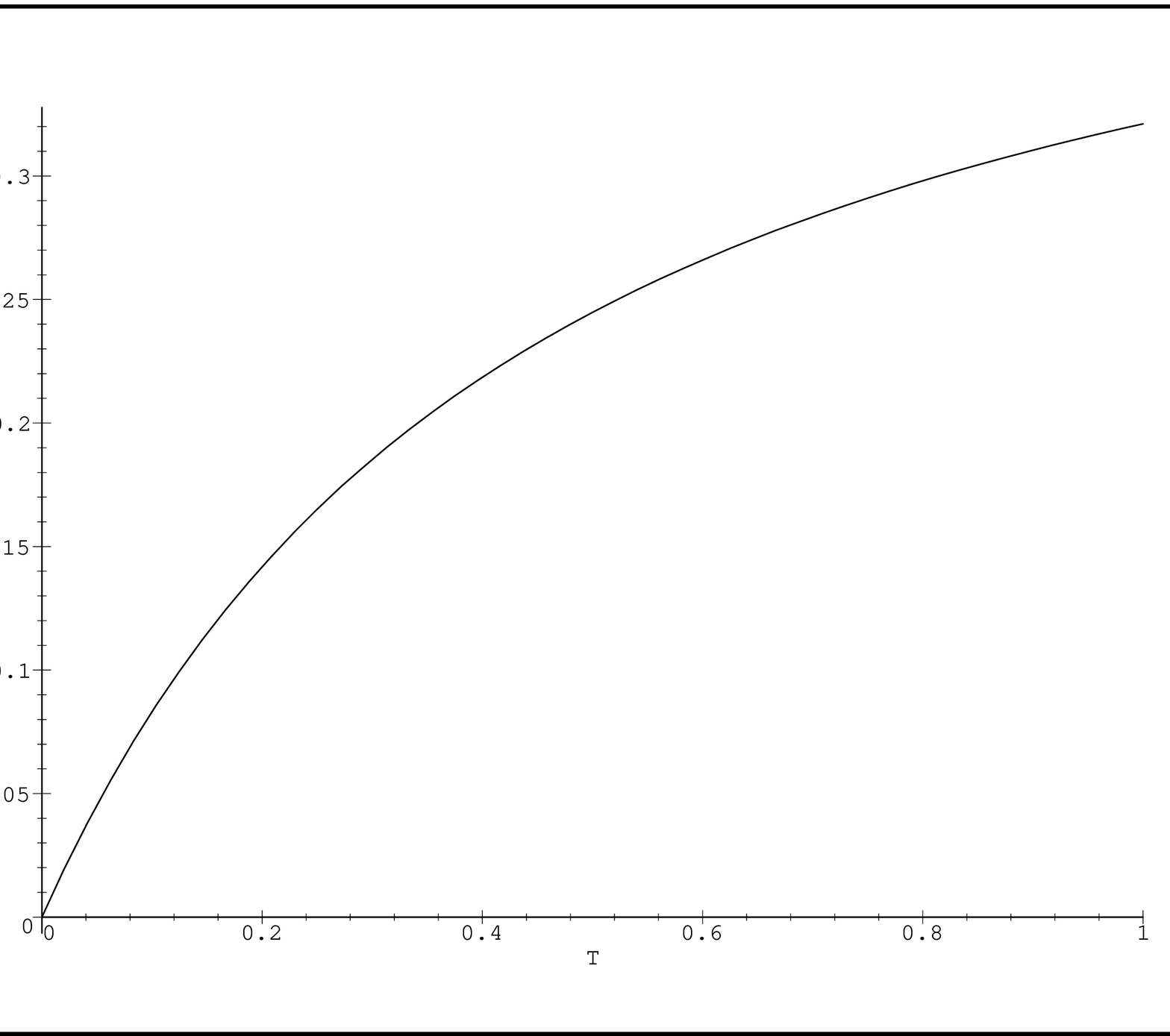}}
\vskip 3ex
\begin{center}
{\small{Fig. 3c}\\
 The total angle in the critical case for the same parameters. 
}
\end{center}

%%%%%%%%%%%%%%%%%%%%%%%%%%%%%  end critical  %%%%%%%%%%%%%%%%%%%%%%%%%%%%%%%%%
%%%%%%%%%%%%%%%%%%%%%%%%%%%%%%  u const  %%%%%%%%%%%%%%%%%%%%%%%%%%%%%%%
\vskip 1ex
\centerline{
\epsfxsize=180pt
\epsfbox{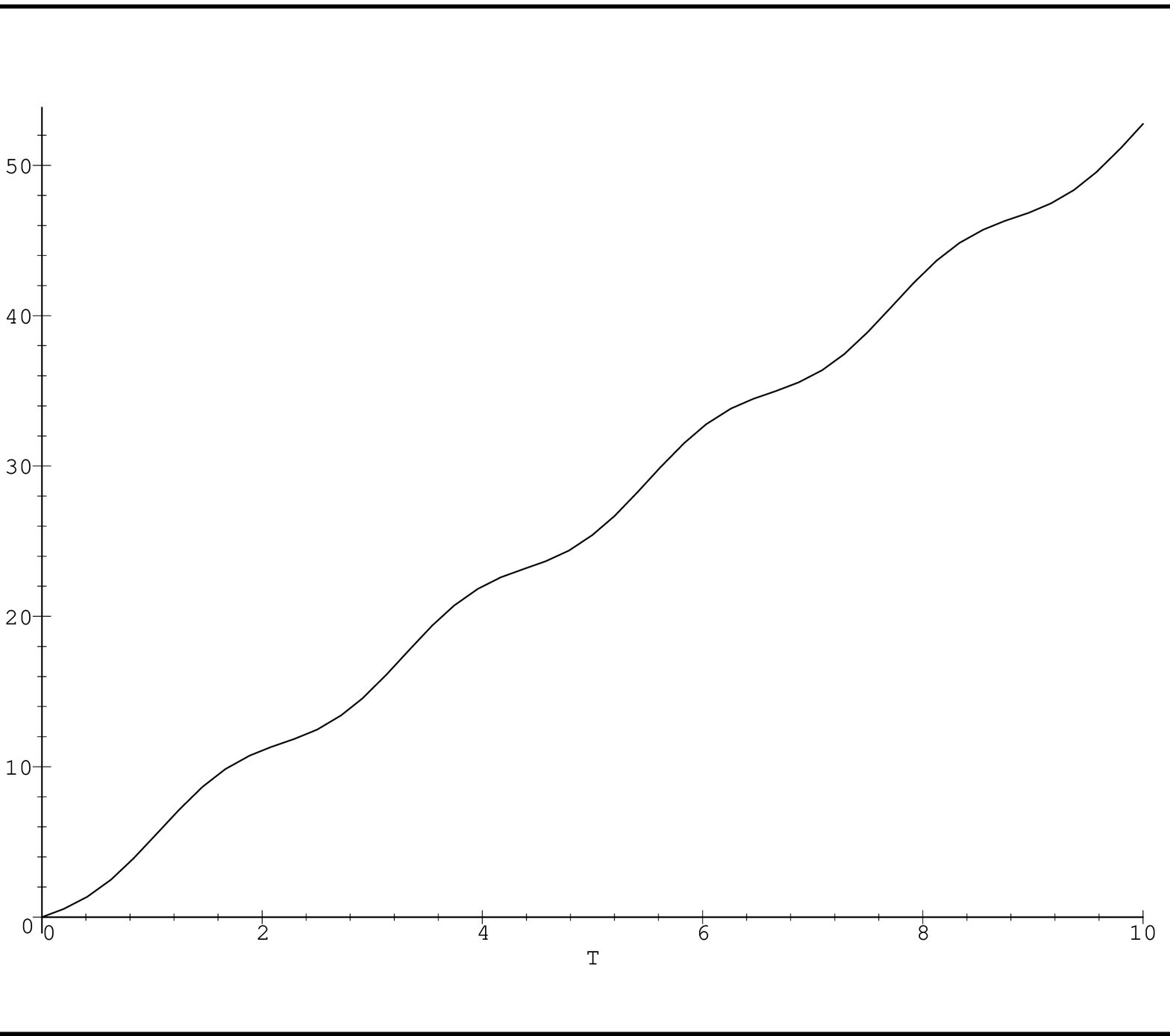}}
\vskip 3ex
\begin{center}
{\small{Fig. 1a'}\\
 The dynamical angle in the underdamped case for the same $\omega _{0}$, $\beta$
parameters as in Fig. 1a and $\gamma =0$. 
}
\end{center}

\vskip 1ex
\centerline{
\epsfxsize=180pt
\epsfbox{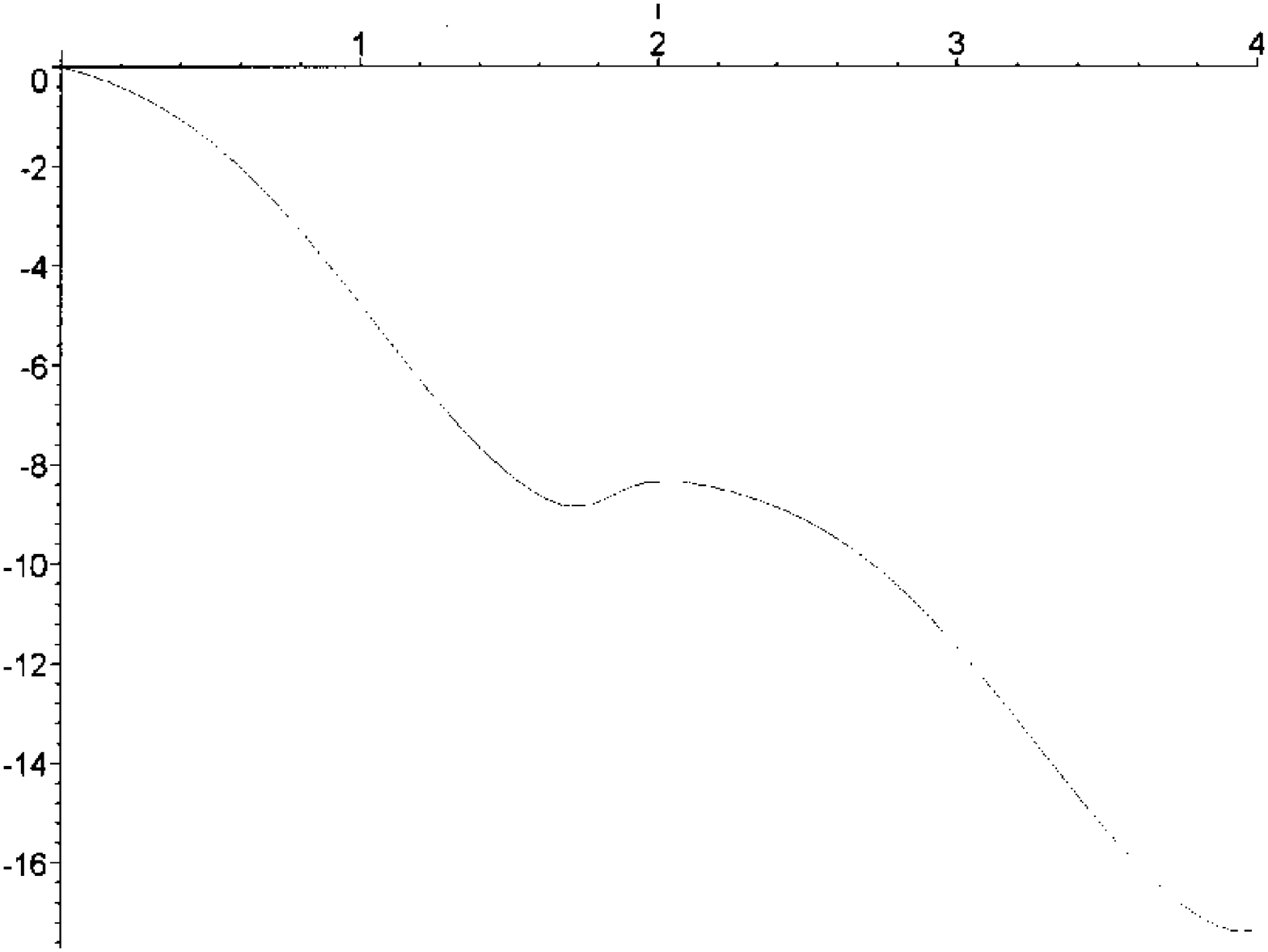}}
\vskip 3ex
\begin{center}
{\small{Fig. 1b'}\\
 The geometrical angle in the underdamped case for the same $\omega _{0}$, 
$\beta$ parameters and $\gamma =0$. 
}
\end{center}

\vskip 1ex
\centerline{
\epsfxsize=180pt
\epsfbox{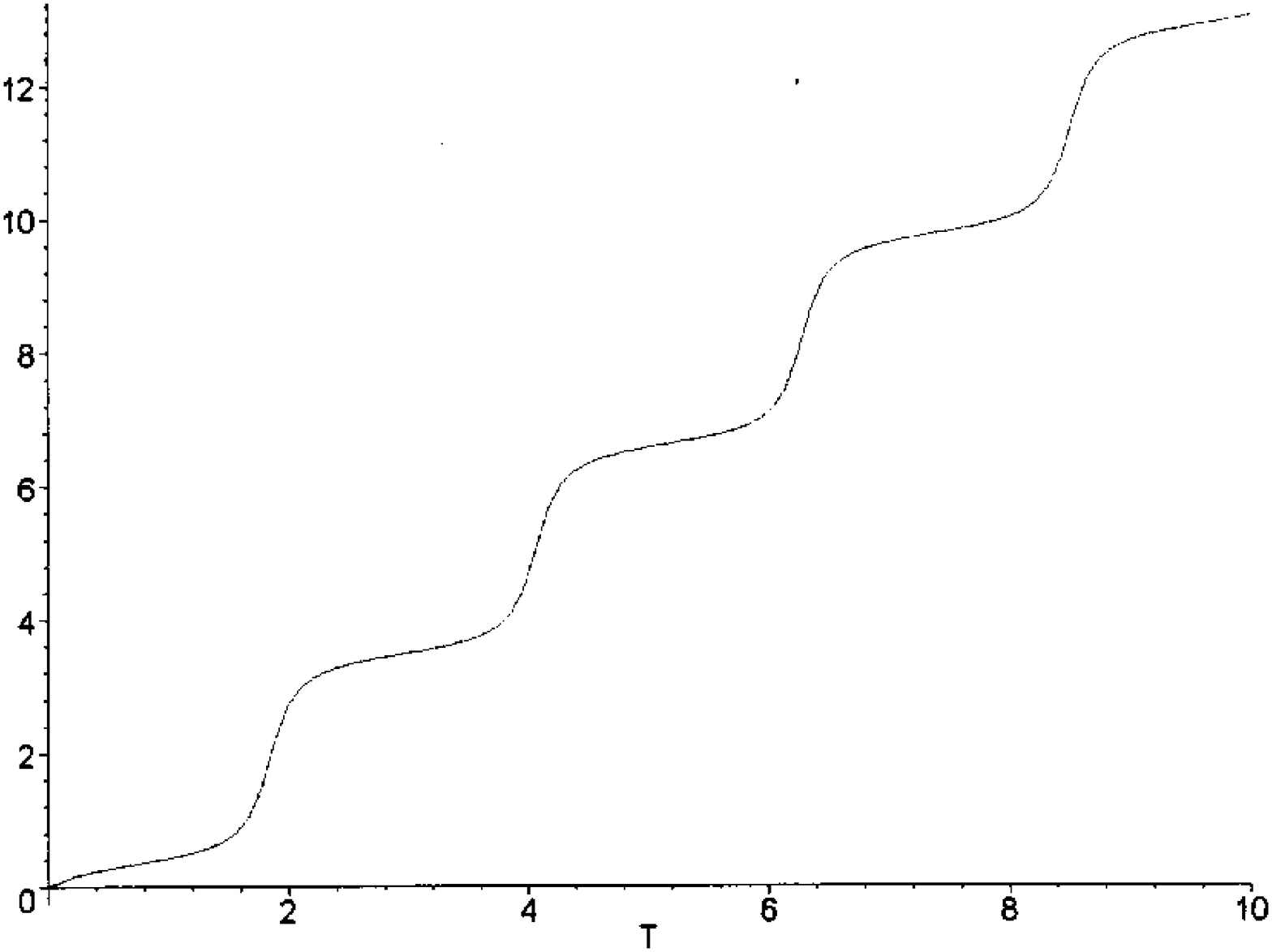}}
\vskip 3ex
\begin{center}
{\small{Fig. 1c'}\\
 The total angle in the underdamped case for the same $\omega _{0}$, $\beta$
parameters and $\gamma =0$. 
}
\end{center}

%%%%%%%%%%%%%%%%%%%%%%%%%%%%%%%% o const %%%%%%%%%%%%%%%%%%%%%%%%%%%%%%%
\vskip 1ex
\centerline{
\epsfxsize=180pt
\epsfbox{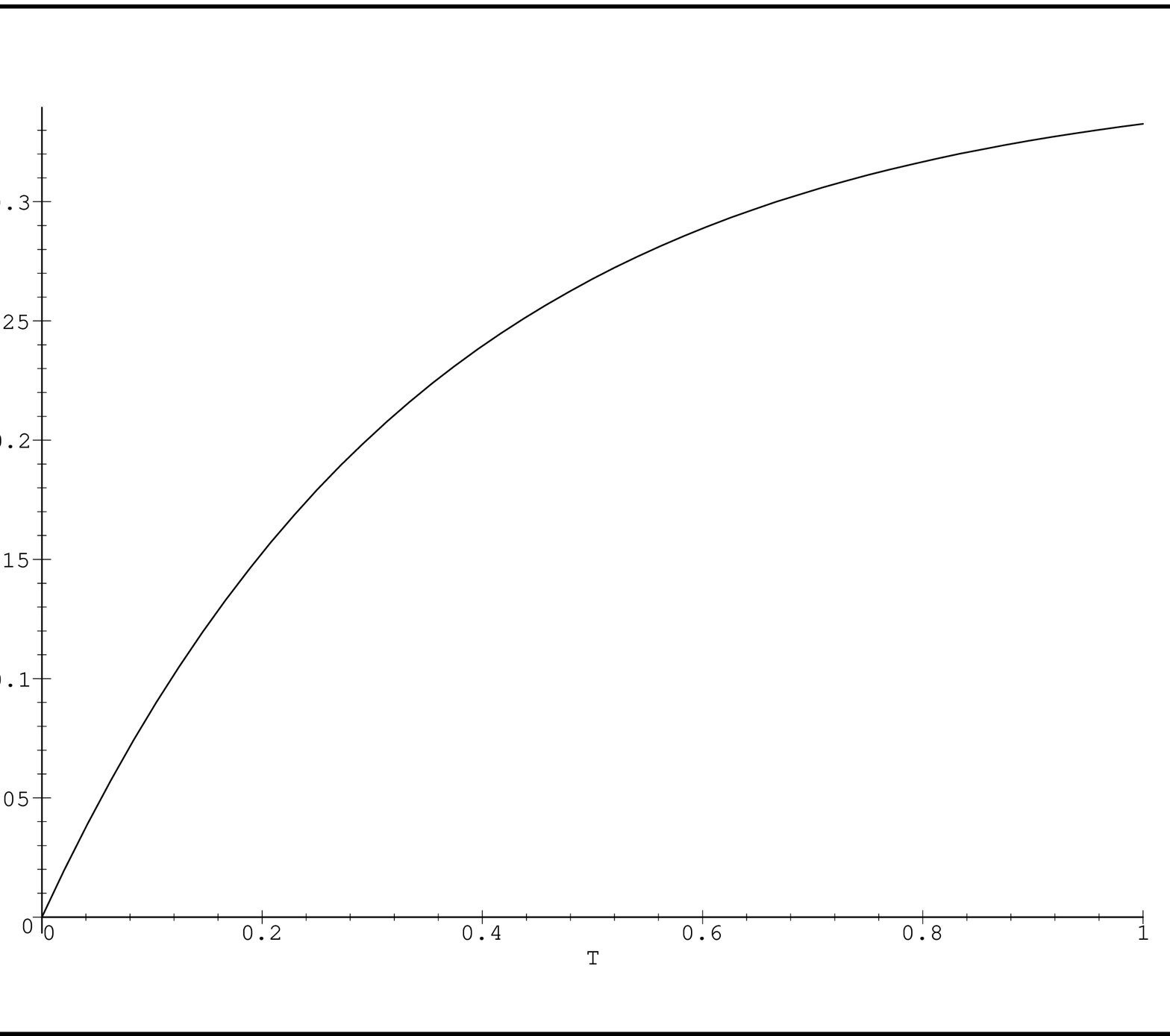}}   %adocons9 | adocons9.eps | April 12
\vskip 3ex
\begin{center}
{\small{Fig. 2a'}\\
 The dynamical angle in the overdamped case for the same $\omega _{0}$, $\beta$
parameters as in Fig. 2a and $\gamma =0$. 
}
\end{center}

\vskip 1ex
\centerline{
\epsfxsize=180pt
\epsfbox{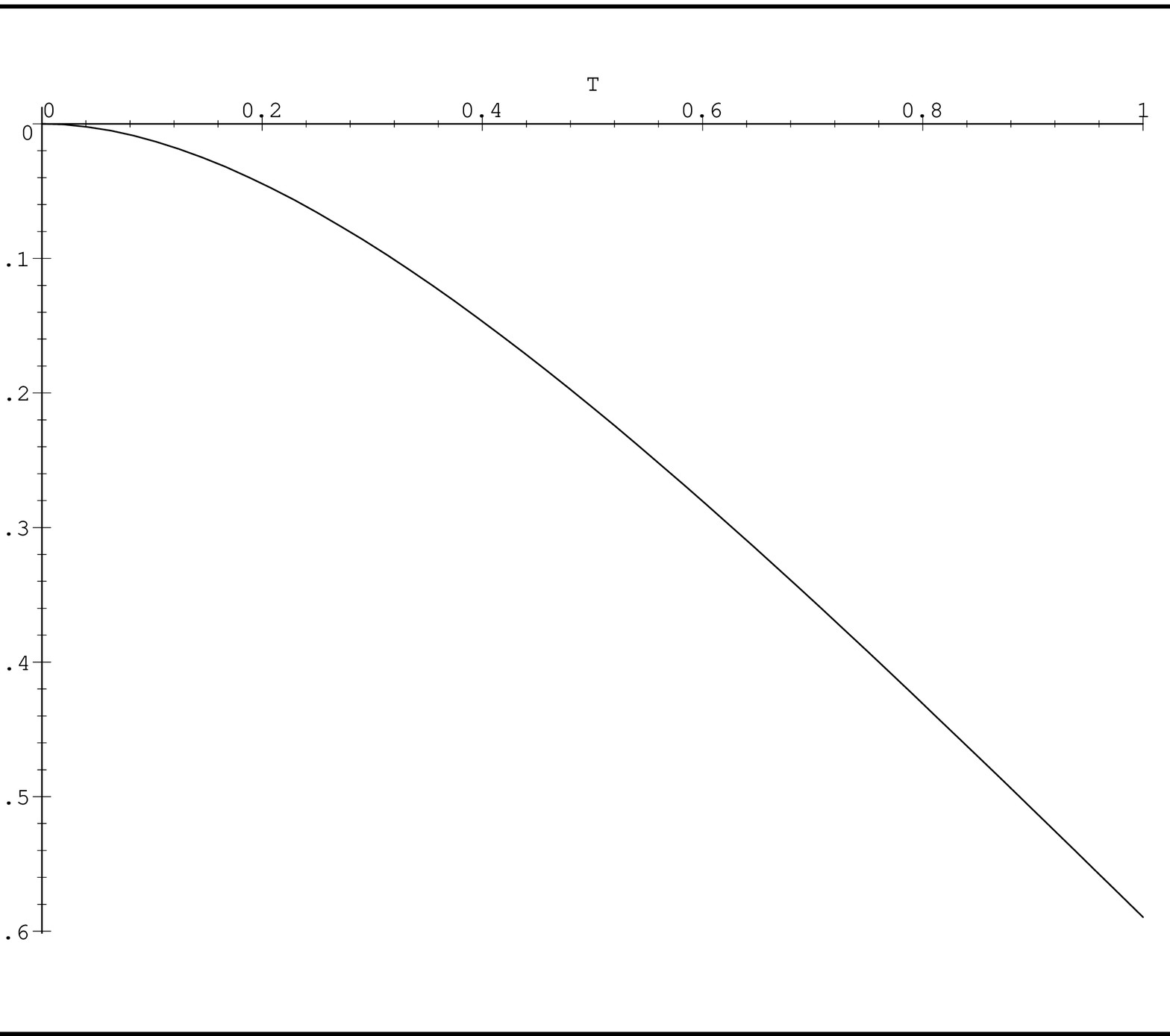}}   %agocons9
\vskip 3ex
\begin{center}
{\small{Fig. 2b'}\\
 The geometrical angle in the overdamped case for the same $\omega _{0}$, $\beta$
parameters and $\gamma =0$. 
}
\end{center}

\vskip 1ex
\centerline{
\epsfxsize=180pt
\epsfbox{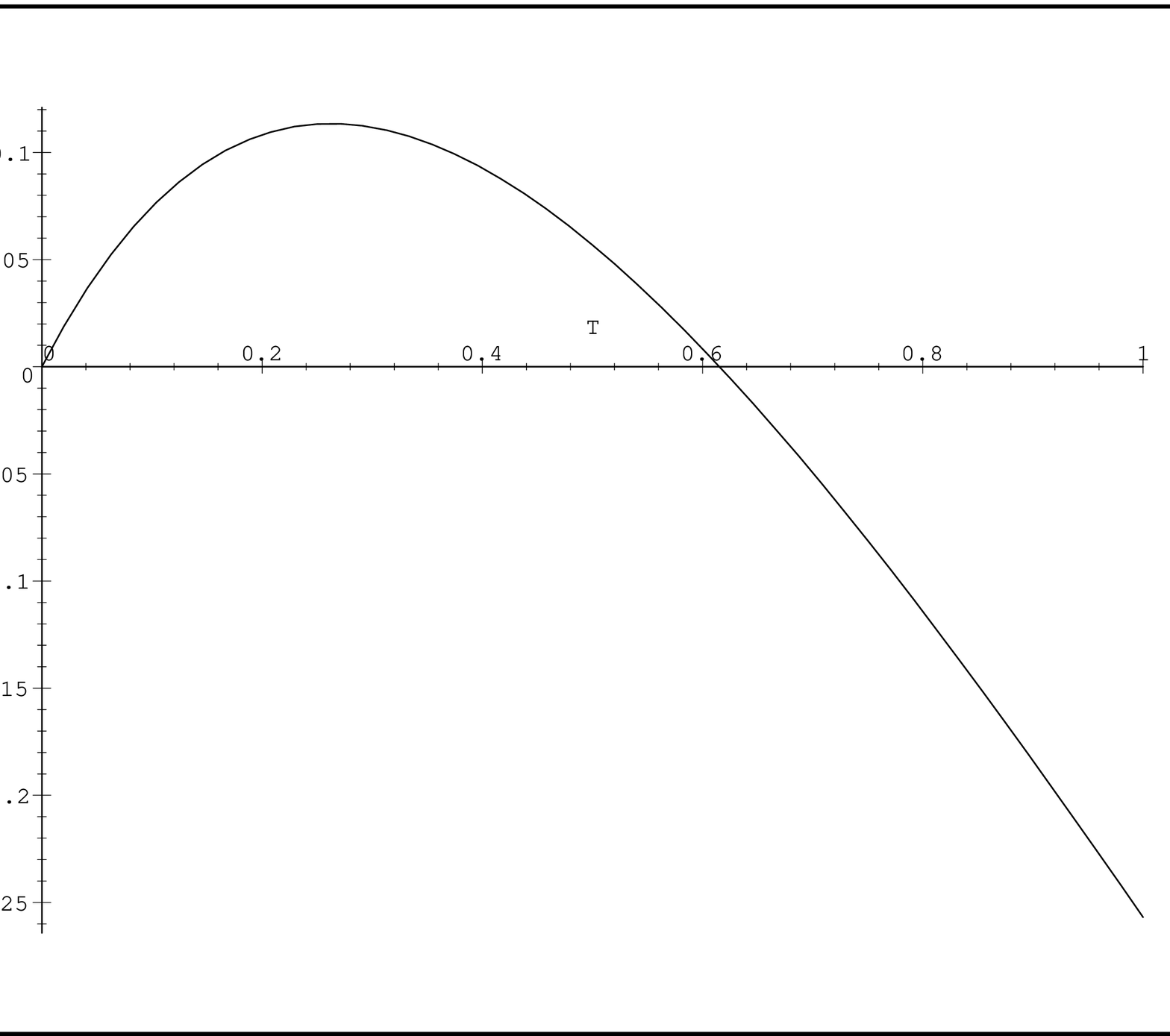}}
\vskip 3ex
\begin{center}
{\small{Fig. 2c'}\\
 The total angle in the overdamped case for the same $\omega _{0}$, $\beta$
parameters and $\gamma =0$. 
}
\end{center}
%%%%%%%%%%%%%%%%%%%%%  c const  %%%%%%%%%%%%%%%%%%%%%%%%%%%%%%%%%

\vskip 1ex
\centerline{
\epsfxsize=180pt
\epsfbox{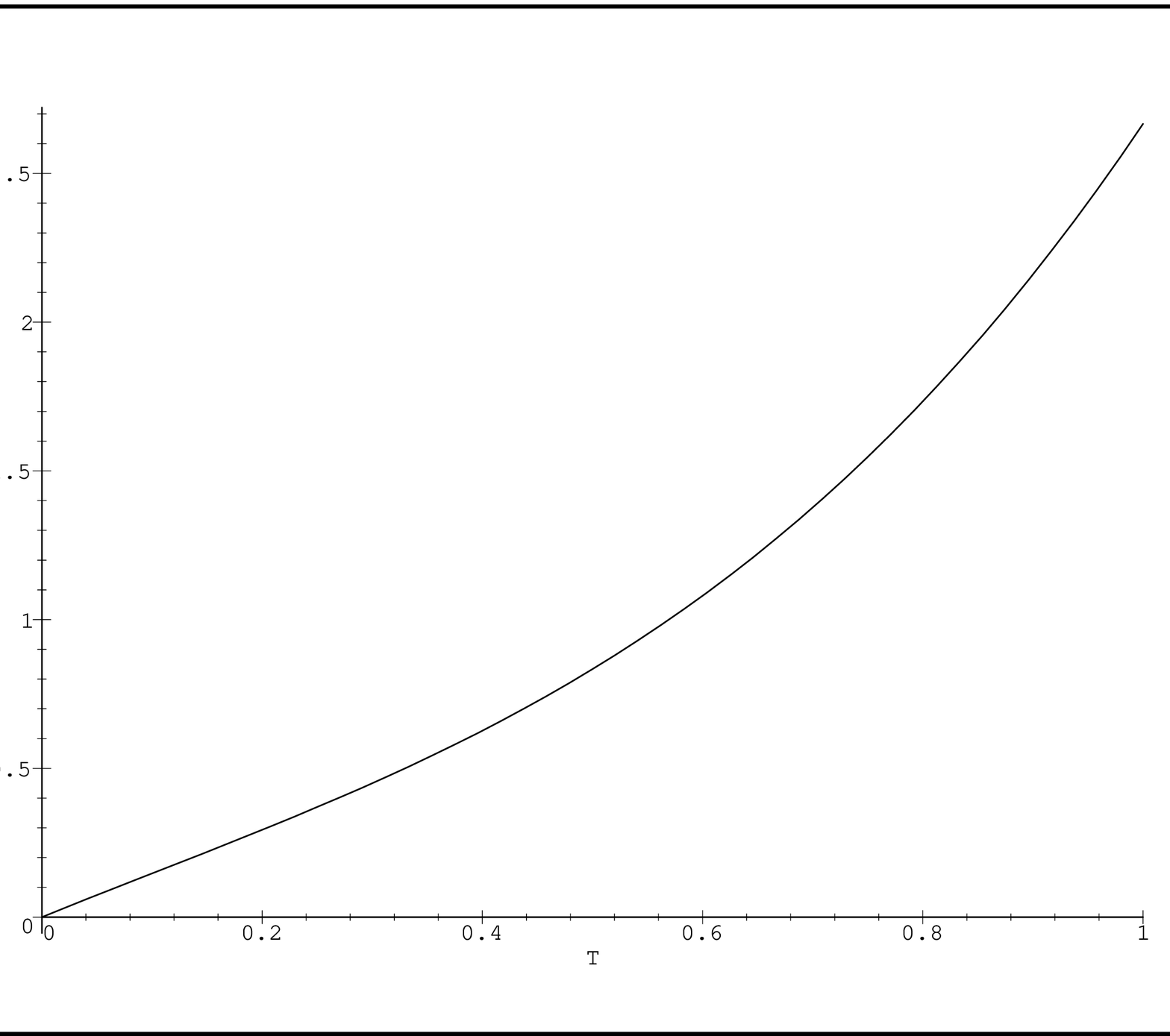}}  %adccons9.eps
\vskip 3ex
\begin{center}
{\small{Fig. 3a'}\\
 The dynamical angle in the critical case for $\omega _{0}=\beta=1$
and $\gamma =0$. 
}
\end{center}

\vskip 1ex
\centerline{
\epsfxsize=180pt
\epsfbox{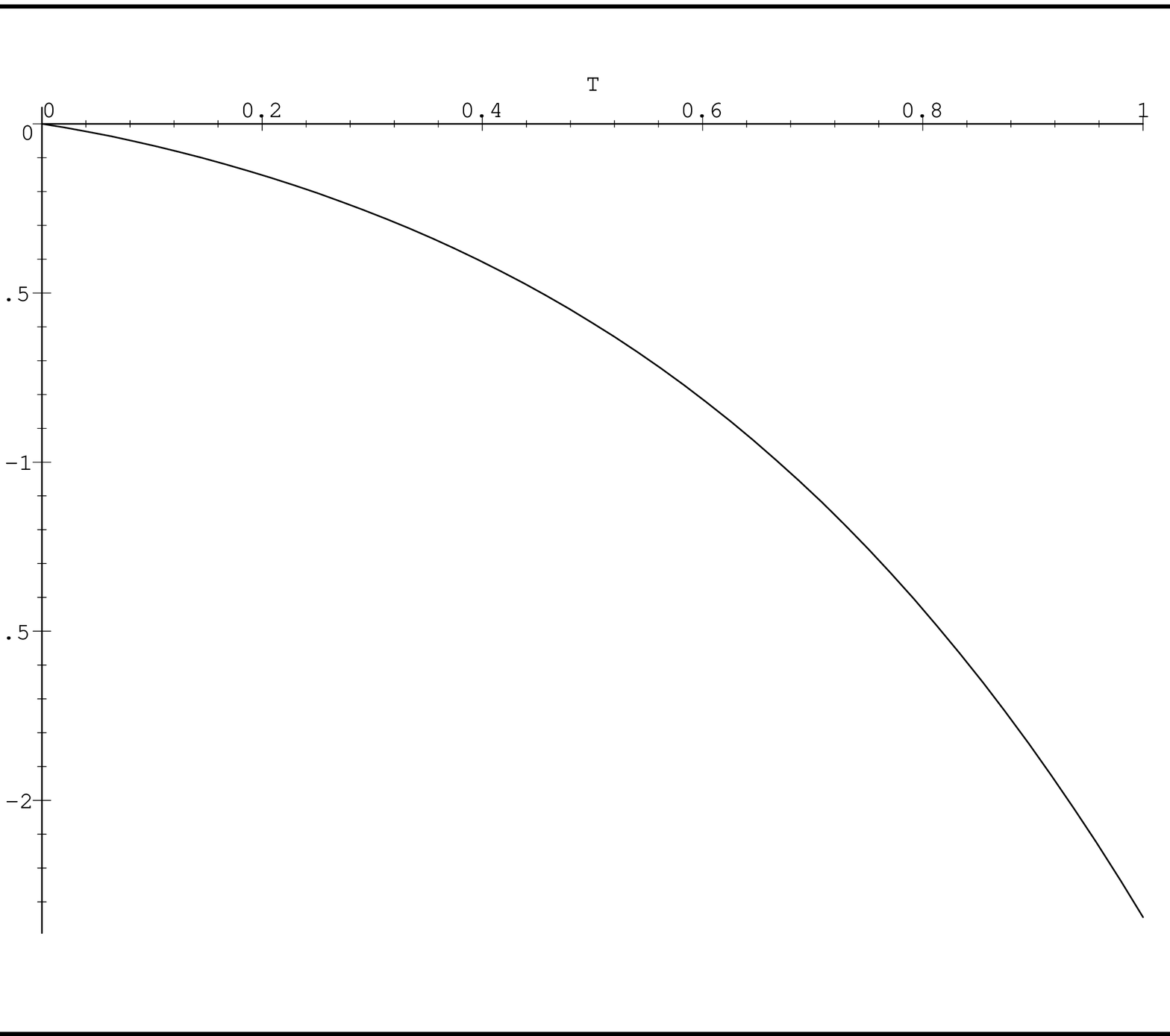}}  %agccons9.eps
\vskip 3ex
\begin{center}
{\small{Fig. 3b'}\\
 The geometrical angle in the critical case for $\omega _{0}=\beta=1$
and $\gamma =0$. 
}
\end{center}

\vskip 1ex
\centerline{
\epsfxsize=180pt
\epsfbox{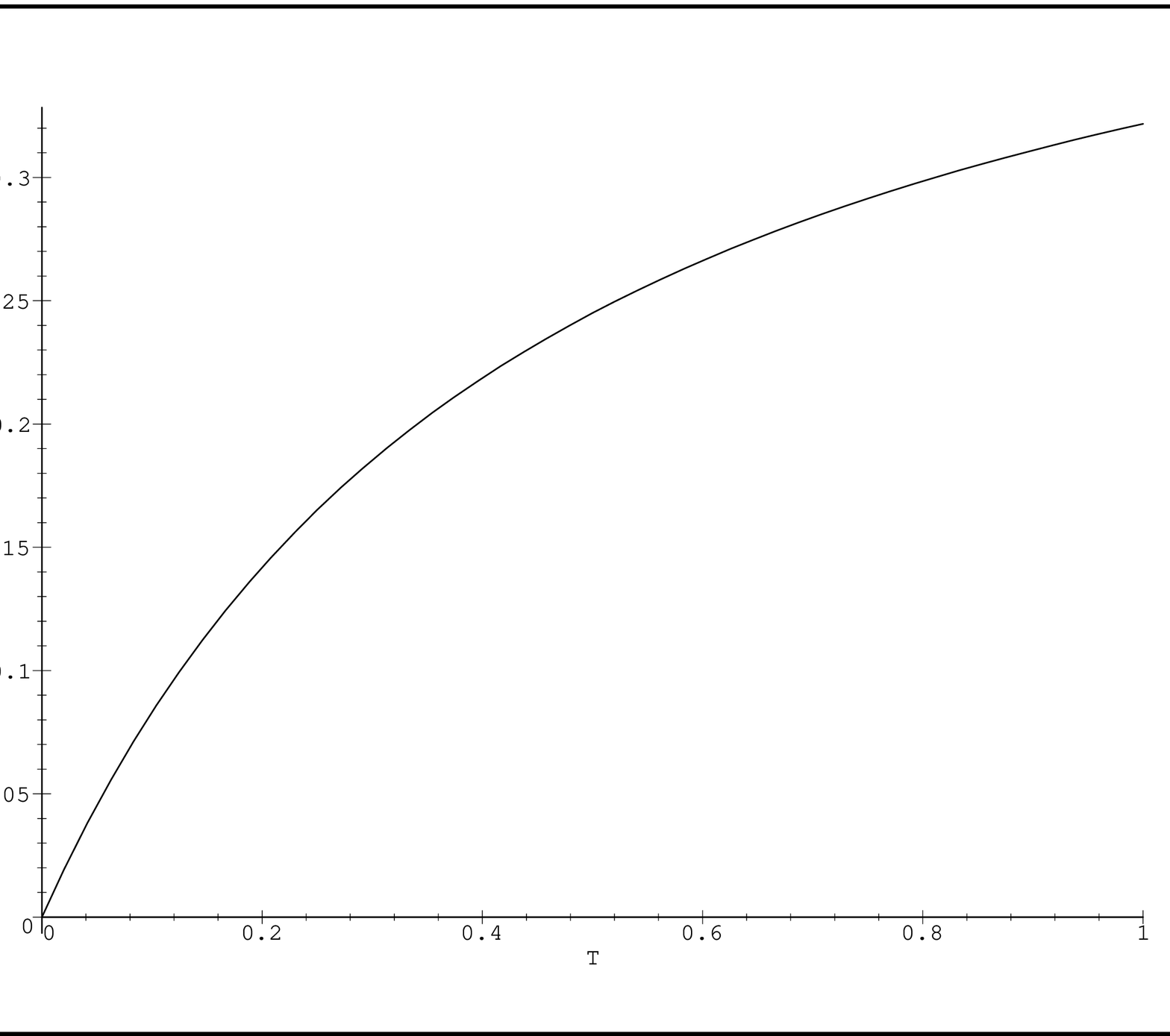}}   %atccons9.eps
\vskip 3ex
\begin{center}
{\small{Fig. 3c'}\\
 The total angle in the critical case for $\omega _{0}=\beta=1$
and $\gamma =0$. 
}
\end{center}
%%%%%%%%%%%%%%%%%%%%%%%%%%%%%%%%%%%%%%%%%%%%%%%%%%%%%%%%%%%%%%%%%%%%%%%%%%%%%
%%%%%%%%%%%%%%%%%%%%%%%%%%%%%%%%%%%%%%%%%%%%%%%%%%%%%%%%%%%%%%%%%%%%%%%

\end{document}